\documentclass[
 reprint,         
 superscriptaddress, 
 amsmath,amssymb, 
 aps,             
 prd,             
 nofootinbib,     
 floatfix         
]{revtex4-2}

\usepackage{graphicx}
\usepackage{subfigure}
\usepackage{caption}
\usepackage{dcolumn}
\usepackage{bm}
\usepackage{physics}
\usepackage[normalem]{ulem}   
\usepackage{xcolor}           
\usepackage{hyperref}         

\hypersetup{
  pdfauthor   = {Breno Agatão, Pedro Brandão, et al.},
  pdftitle    = {Correlation function for n Ds1(2460) and n Ds1(2536)},
  colorlinks  = true,
  linkcolor   = blue,
  citecolor   = blue,
  urlcolor    = blue
}

\newcommand{\USP}{Universidade de São Paulo, Instituto de Física, 05389‑970, São Paulo, Brazil}
\newcommand{\IFIC}{Departamento de Física Teórica and IFIC, Centro Mixto Universidad de Valencia‑CSIC, Institutos de Investigación de Paterna, 46071 Valencia, Spain}
\newcommand{\UFBA}{Universidade Federal da Bahia, Instituto de Física, 40170-115, Bahia, Brazil}
\newcommand{\UNIFESP}{Universidade Federal de São Paulo, 01302-907, São Paulo, Brazil}
\newcommand{\GKLNPT}{Guangxi Key Laboratory of Nuclear Physics and Technology, Guangxi Normal University, Guilin 541004, China}
\begin{document}

\title{Correlation functions for \texorpdfstring{$n\,\bar{D}_{s1}(2460)$ and $n\,\bar{D}_{s1}(2536)$}{n Ds1(2460) and n Ds1(2536)}}

\author{Breno Agatão}
\email{bgarcia@if.usp.br}
\affiliation{\USP}
\affiliation{\IFIC}

\author{Pedro Brandão}
\email{pedro.brandao@ufba.br}
\affiliation{\UFBA}
\affiliation{\IFIC}

\author{A. Martínez Torres}
\email{amartine@if.usp.br}
\affiliation{\USP}

\author{K. P. Khemchandani}
\email{kanchan.khemchandani@unifesp.br}
\affiliation{\UNIFESP}

\author{Luciano M. Abreu}
\email{luciano.abreu@ufba.br}
\affiliation{\UFBA}

\author{E. Oset}
\email{oset@ific.uv.es}
\affiliation{\IFIC}
\affiliation{\GKLNPT}

\date{\today}

\begin{abstract}
We study the interaction of the  $n \bar {D}_{s1}(2460)$ and $n \bar {D}_{s1}(2536)$ systems from the perspective that the $D_{s1}(2460)$ and $D_{s1}(2536)$ states are molecular states of  $K D^*$ and $K^*D$, respectively. We use an improved version of the fixed center approximation to the Faddeev equations which fulfills exact elastic unitarity at the threshold of the systems. We predict the existence of resonant states below threshold and also determine the scattering length, effective range and correlation functions of these systems, showing that the results contain important information to test the assumed molecular nature of the two $D_{s1}$ states.
\end{abstract}

\maketitle


\section{Introduction}

Originally developed to deal with astronomical issues~\cite{Brown54,Brown56}, femtoscopic correlation functions have recently emerged as a very useful tool to learn about hadron interactions, by measuring pairs of particles in ultrarelativistic $pp$, $pA$, $AA$ reactions~\cite{Fabbietti21,ALICE17,ALICE19,ALICE19_2,ALICE19_3,ALICE20,ALICE20_2,ALICE20_3,ALICE21,ALICE21_2,ALICE22,STAR15,STAR19,Feijoo25,ALICE23}. These measurements give us the opportunity to learn about the interaction of pairs of particles which often cannot be obtained from scattering experiments. Theoretical developments have followed, proposing methods to evaluate them~\cite{KOONIN77,Lednicky81,Pratt86,Pratt87,Pratt90,Bauer92}, making predictions for different systems or comparing with data~\cite{Morita:2014kza,Ohnishi:2016elb,Platonova:2010wjt,Hatsuda:2017uxk,Mihaylov:2018rva,Haidenbauer:2018jvl,Morita:2019rph,Kamiya:2019uiw,Kamiya:2021hdb,Kamiya:2022thy,Vidana:2023olz,Liu:2023uly,Albaladejo:2023pzq,Torres-Rincon:2023qll,Sarti:2023wlg,Molina:2023oeu,Molina:2023jov,Liu:2024nac,Feijoo:2024bvn,Ikeno:2023ojl,Albaladejo:2023wmv,Feijoo:2023sfe,Khemchandani:2023xup,Abreu:2024qqo,Abreu:2025jqy,Encarnacion:2025luc,Ramos:2025ibe}. A recent paper digs into the relevant issue of the relationship of correlation functions and usual production experiments that measure invariant mass distributions in the same reactions~\cite{Albaladejo:2024lam}. Similarly, the simultaneous consideration of spectroscopy and the correlation data opens new ways for a steady progress in our understanding of hadron dynamics~\cite{Feijoo:2024qqg}. 
\par The success in the study of the two body correlation functions has opened the doors to the study of three body correlation functions~\cite{DelGrande:2021mju,ALICE:2022boj,ALICE:2023gxp,ALICE:2023bny,Garrido:2024pwi,Garrido:2025lar}. But out of these, we find particularly useful the study of the correlation functions of a stable particle and a resonance, which is just emerging as a new promising field. The first of such experiments is underway~\cite{SersknytePrivate2024}, looking at the correlation function of a proton and the $f_1(1285)$ axial vector resonance.
\par The $f_1(1285)$, together with related axial vector meson resonances, appears in a natural way as a consequence of the interaction of pseudoscalar mesons with vector mesons in S-wave in the context of the chiral unitary approach~\cite{Lutz:2003fm,Roca:2005nm,Garcia-Recio:2010enl,Zhou:2014ila,Geng:2015yta,Lu:2016nlp}. Concretely, the $f_1(1285)$ appears from the interaction of the $K^*\bar{K}$, $\bar{K}^*K$ systems, via the combination $K^*\bar{K} - \bar{K}^*K$ in isospin $I=0$.
\par The measurement of such correlation functions can shed light on the structure of these resonances, where a current debate is underway about their nature as ordinary states ($q\bar{q}$ for mesons or $qqq$ for baryons), tetraquarks or pentaquarks, molecular states, etc.~\cite{Esposito:2016noz,Guo:2017jvc,Olsen:2017bmm,Lebed:2016hpi,Chen:2016qju,Liu:2019zoy,Brambilla:2019esw,Ali:2017jda,Karliner:2017qhf}. A step in this direction was given in Ref.~\cite{Encarnacion:2025lyf}, where the correlation function of $pf_1(1285)$ was calculated from the perspective of the $f_1(1285)$ as a molecular state, and it was found that differences appeared in the correlation functions depending on the structure assumed for the $f_1(1285)$.
\par In Ref.~\cite{Encarnacion:2025lyf} the fixed center approximation (FCA)~\cite{Foldy:1945zz,Brueckner:1953zz,Brueckner:1953zza,Chand:1962ec,Barrett:1999cw,Deloff:1999gc,Kamalov:2000iy,MartinezTorres:2020hus,Roca:2010tf,Malabarba:2024hlv} was used to study the $pf_1(1285)$, assuming the $K^*\bar{K}$, $\bar{K}^*K$ system as the cluster and the proton as the external particle interacting with the constituents of the cluster. The method not only led to finding a bound state about $40$~MeV below the $pf_1(1285)$ threshold, but also predicted the corresponding correlation functions. However, it was found that the FCA formula did not exactly fulfill elastic unitarity around the $pf_1(1285)$ threshold, but it could be re-established by multiplying the FCA amplitude by a factor close to unity, which allowed to properly evaluate the correlation functions.   
\par Following this work, the issue of elastic unitarity was retaken in Ref.~\cite{Ikeno:2025bsx} and the standard FCA was improved to exactly fulfill elastic unitarity. In Ref.~\cite{Ikeno:2025bsx} the correlation function of the $n\bar{D}_{s0}(2317)$ interaction was studied assuming the $D_{s0}(2317)$ to be a molecular state formed from $DK$, $D_s\eta$ interactions~\cite{MartinezTorres:2014kpc,vanBeveren:2003kd,Barnes:2003dj,Chen:2004dy,Kolomeitsev:2003ac,Gamermann:2006nm,Guo:2006rp,Yang:2021tvc,Liu:2022dmm}, mostly made from a $DK$ system in $I=0$. By elastic unitarity we mean that the Quantum Mechanics scattering amplitude for $n\bar{D}_{s0}(2317)$ scattering must behave, close to threshold, as 
\begin{equation}
    (f^{QM})^{-1}\approx - \frac{1}{a}+\frac{1}{2}r_0k^2-ik,
\end{equation}
\noindent with $k$ being the modulus of the momentum in the $n\bar{D}_{s0}$ rest frame, $a$ the scattering length and $r_0$ the effective range. Although $a,\ r_0$ depend on the dynamics of the process, the $-ik$ term is independent of the dynamics and is what makes $f^{QM}$ to fulfill unitarity, relating $\text{Im}\{f^{QM}\}$ to the cross section of the reaction. The improvement to the FCA formula came from considering elastic propagation of the $n$ and the $\bar{D}_{s0}(2317)$ as a cluster different than the scattering of the $n$ from the $\bar K$ to the $\bar D$ in the cluster. The procedure followed the steps used in nuclear physics problems, where the interaction of an external particle  with a nucleus proceeds via the evaluation of an optical potential, followed by the solution of the Lippmann-Schwinger equation (Schrödinger equation) with this potential~\cite{Ericson:1988gk,Seki:1983sh,Nieves:1993ev,Brown:1975di}. This was the guiding principle to construct the fully unitary formula for the interaction of the external particle with the cluster in Ref.~\cite{Ikeno:2025bsx}.
\par In the present work, we continue the line initiated in Refs.~\cite{Encarnacion:2025lyf,Ikeno:2025bsx} and we study the interaction of the $n \bar{D}_{s1}(2460)$ and $n \bar{D}_{s1}(2536)$ systems. The $D_{s1}(2460)$ and $D_{s1}(2536)$ states have also been advocated as molecular states, mostly formed respectively from $D^*K$ and $DK^*$, ~\cite{Kolomeitsev:2003ac,Hofmann:2003je,Gamermann:2007fi,Cleven:2010aw,Cleven:2014oka} and lattice QCD calculations also support this picture~\cite{MartinezTorres:2014kpc,Bali:2017pdv,Song:2022yvz}. The choice of $n$ and $\bar{D}_{s1}$ is because we avoid the Coulomb interaction and we have the $N\bar{K}$, $N\bar{K}^*$ interactions, which are attractive and facilitate the formation of states below threshold. We use the unitary formalism of Ref.~\cite{Ikeno:2025bsx}, but are able to simplify considerably the analytical form of the final amplitude, which allows one to see clearly the exact unitarity of the formula.
\par From the physical point of view, we find bound states below the $n\bar{D}_{s1}$ thresholds and evaluate the correlation function for the two systems, plus the scattering length and effective range.

\section{Formalism}
\subsection{The \texorpdfstring{$n\,\bar{D}_{s1}(2460)$}{n Ds1(2460)} system}

In the $C=1$, $S=1$ sector, the $D^*K$ interaction dynamically generates $D_{s1}(2460)$, a state with $I=0$ and $J^P=1^+$ \cite{Gamermann:2007fi}, which, using the isospin phase convention $D^*= \begin{Bmatrix}D^{*+} \\ - D^{*0} \end{Bmatrix}$ and $K= \begin{Bmatrix}K^{+} \\  K^{0} \end{Bmatrix}$, can be written as
\begin{align}
    \ket{D_{s1}(2460)} = \frac{1}{\sqrt{2}}\left( \ket{D^{*+}K^{0}} + \ket{D^{*0}K^{+}} \right).
    \label{eq:Ds12460}
\end{align}
\par As previously mentioned, we consider the $n\bar{D}_{s1}(2460)$ system instead of $pD_{s1}(2460)$ since, not only we avoid dealing with the Coulomb interaction when determining the corresponding correlation function, but also the $n\bar{K}$ interaction present in one of the two-body subsystems is known to be attractive and dynamically generates the $\Lambda(1405)$~\cite{Oller:2000fj,Jido:2003cb,MartinezTorres:2012yi,Hyodo:2011ur,Oset:2012gi}.
\par
\begin{figure}[htbp]
\begin{center}
\includegraphics[scale=0.55]{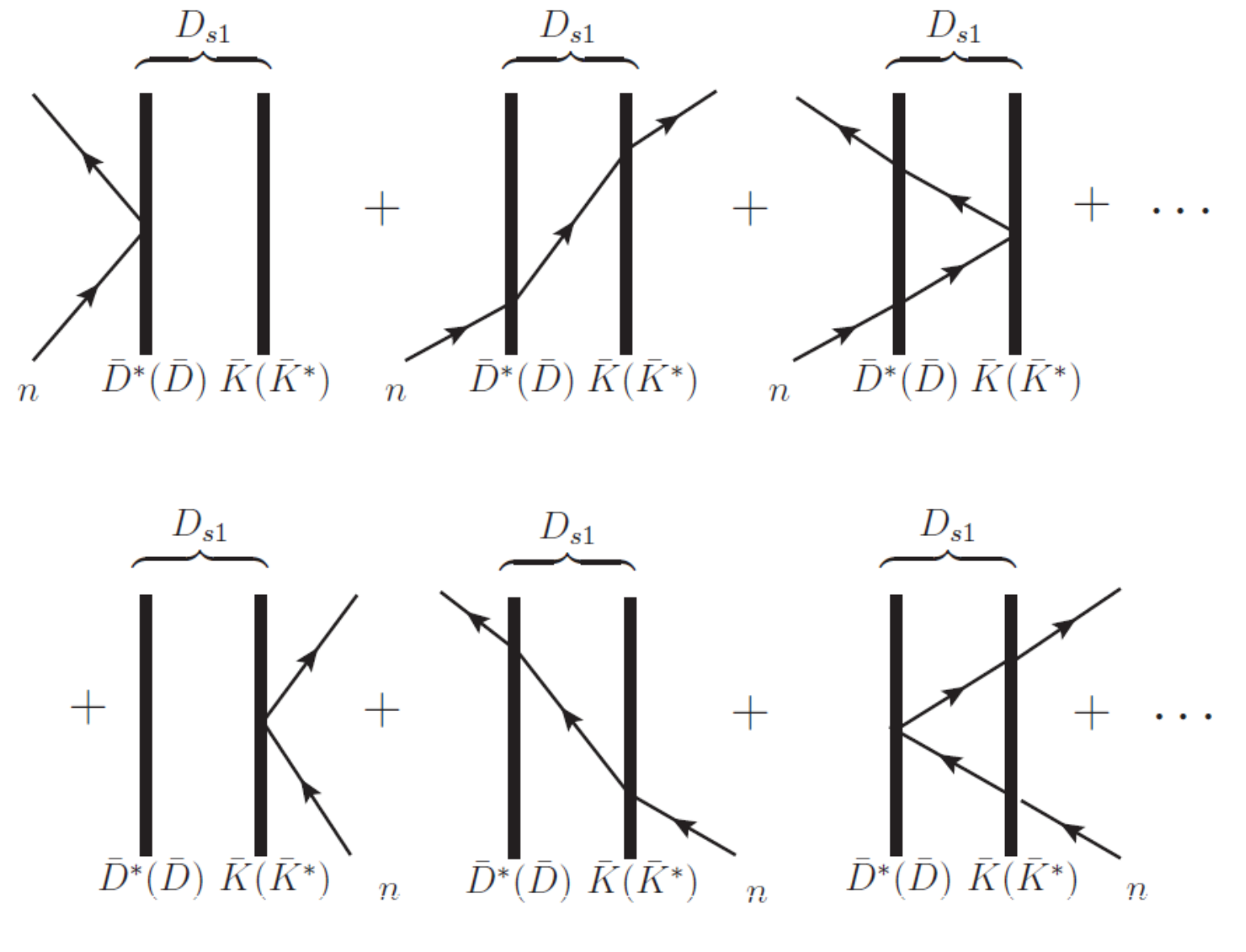}
\caption{Diagrams contributing in the FCA to the $n(\bar{D}^{*}\bar{K})_{\text{cluster}}$, $n(\bar{D}\bar{K}^*)_{\text{cluster}}$ interaction.}
\label{fig:FCA_old}
\end{center}
\end{figure}

 Within the framework of the FCA \cite{Encarnacion:2025lyf,Roca:2010tf,Malabarba:2024hlv}, we consider the $\bar{D}^*\bar{K}$ to cluster as $\bar{D}_{s1}(2460)$ and the $n$ to rescatter successively with either the $\bar{D}^*$ or $\bar{K}$ forming the mentioned cluster (see Fig.~\ref{fig:FCA_old}). This means that we need to describe both the $n\bar{D}^*$ and $n\bar{K}$ interactions (we refer the reader to Appendix~\ref{app:amp} for more details on the calculation of the corresponding two-body $t$-matrices). Since the cluster has $I=0$, the isospin combination for $n \bar{D}_{s1}(2460)$ is given by 
\begin{align}
    \ket{I_{n\bar{D}_{s1}}=1/2,I_{3}=-1/2} & = \ket{I_N=1/2,I_3=-1/2}\nonumber\\ & \qquad \otimes\ket{I_{\bar{D}_{s1}}=0,I_3=0},
    \label{eq:nDs1}
\end{align}
where
\begin{align}
    &\ket{I_{\bar{D}_{s1}}=0,I_3=0} = \frac{1}{\sqrt{2}}\ket{I_{\bar{D}^*}=1/2,I_3=1/2}\nonumber\\& \quad\otimes \ket{I_{\bar{K}}=1/2,I_3=-1/2} - \frac{1}{\sqrt{2}}\ket{I_{\bar{D}^*}=1/2,I_3=-1/2}\nonumber\\& \quad \otimes \ket{I_{\bar{K}}=1/2,I_3=1/2},
    \label{eq:Ds1_iso}
\end{align}
and we can use Eqs.~\eqref{eq:nDs1} and \eqref{eq:Ds1_iso} to determine the amplitudes describing the interaction between the $n\bar{D}^*/n\bar{K}$ systems. To do this, we just need to combine either the ket $\ket{I_N=1/2,I_3=-1/2}$ with the kets $\ket{I_{\bar{D}^*},I_3}$ in Eqs.~\eqref{eq:nDs1} and \eqref{eq:Ds1_iso} in terms of kets $\ket{I_{N\bar{D}^*},I_3}$ or $\ket{I_N=1/2,I_3=-1/2}$ with the kets $\ket{I_{\bar{K}},I_3}$ in terms of $\ket{I_{N\bar{K}},I_3}$ and calculate the transition amplitude
\begin{equation}
\bra{I_{n\bar{D}_{s1}}=1/2,I_{3}=-1/2}t\ket{I_{n\bar{D}_{s1}}=1/2,I_{3}=-1/2}
\end{equation}
for these two rearrangements of the $\ket{I_{n\bar{D}_{s1}}=1/2,I_{3}=-1/2}$ state.
\par Following this procedure, and naming $t_1$($t_2$) as the amplitude describing the $n\bar{D}^*$($n\bar{K}$) system, we obtain~\cite{Encarnacion:2025lyf,Roca:2010tf,Xiao:2011rc,Ikeno:2022jbb,Bayar:2023itf}
\begin{align}
    t_1(\sqrt{s_{n\bar{D}^*}}) \equiv t_1 &= \frac{3}{4}t_{n\bar{D}^*}^{I=1} + \frac{1}{4}t_{n\bar{D}^*}^{I=0}, \nonumber \\
    t_2(\sqrt{s_{n\bar{K}}}) \equiv t_2 &= \frac{3}{4}t_{n\bar{K}}^{I=1} + \frac{1}{4}t_{n\bar{K}}^{I=0},
    \label{eq:2body_t}
\end{align}
where $t_{n\bar{D}^*}^{I}$ and $t_{n\bar{K}}^{I}$ represent the two-body $t$-matrices describing, respectively, the interactions $n\bar{D}^*\rightarrow n\bar{D}^*$ and $n\bar{K}\rightarrow n\bar{K}$ in a given isospin $I$. 
To calculate the invariant masses $\sqrt{s_{n\bar{D}^*}}$ and $\sqrt{s_{n\bar{K}}}$, we assume that the binding of the $\bar{D}_{s1}(2460)$ is shared between its constituents, i.e., $\bar{D}^*$ and $\bar{K}$, such that the masses of $\bar{D}^*$, $\tilde{M}_{\bar{D}^*}$, and $\bar{K}$, $\tilde{M}_{\bar{K}}$, forming the cluster of mass $M_C$ are proportional to their respective values, $M_{\bar{D}^*}$ and $M_{\bar{K}}$, when they are free particles. To be more specific, we have $\tilde{M}_{\bar{D}^*}+\tilde{M}_{\bar{K}} = M_C$, with $\tilde{M}_{\bar{D}^*}= \xi M_{\bar{D}^*}$, $\tilde{M}_{\bar{K}}= \xi m_{\bar{K}}$ and $\xi = M_C/(M_{\bar{D}^*}+m_{\bar{K}})$. In this form, the arguments of $t_1$ and $t_2$ are given by 
\begin{align}
    s_{n\bar{D}^*}& = (p_n + p_{\bar{D}^*})^2 = M_N^2 + (\xi M_{\bar{D}^*})^2+ 2\xi M_{\bar{D}^*} q^0, \nonumber \\
    s_{n\bar{K}} &= (p_n + p_{\bar{K}})^2 = M_N^2 + (\xi M_{\bar{K}})^2+ 2\xi M_{\bar{K}} q^0,
\end{align}
\noindent where $q^0=(s-M_N^2-M_C^2)/2M_C$ stands for the energy of the neutron in the rest frame of the cluster and $M_N$ for its mass.
\par Following the formalism in Refs.~\cite{Encarnacion:2025lyf,Roca:2010tf,Malabarba:2024hlv}, we need to multiply $t_1$ and $t_2$ by weighting factors to relate the $S$-matrix of $n$ interacting with the $\bar{D}^*$ and $\bar{K}$ forming $\bar{D}_{s1}(2460)$ and that of $n$ interacting with $\bar{D}_{s1}(2460)$, considering the latter as an elementary particle. This is done by setting 
\begin{equation}
    t_1\rightarrow \tilde{t}_1=\frac{M_C}{M_{\bar{D}^*}}t_1, \qquad t_2\rightarrow \tilde{t}_2=\frac{M_C}{M_{\bar{K}}}t_2.
\end{equation}
\par As mentioned before, to satisfy elastic unitarity and be able to calculate correlation functions with the $T$-matrix determined from the FCA, we need to reformulate the FCA of Ref.~\cite{Encarnacion:2025lyf} to account for the propagation of particle-cluster intermediate states, with the cluster propagating as a whole. To do this, it is convenient to write the $T$-matrix obtained with the FCA in terms of four partition functions: 
 $\tilde{T}_{11}$, $\tilde{T}_{12}$, $\tilde{T}_{21}$ and $\tilde{T}_{22}$. Each partition function $\tilde{T}_{ij}$ considers contributions to the scattering that begins with the neutron colliding with particle $i$ of the cluster and finalizes with the neutron interacting with particle $j$ of the cluster. These partition functions fulfill the coupled equations
 \begin{align}
    \tilde{T}_{11} &= \tilde{t}_1 + \tilde{t}_1 G_0 \tilde{T}_{21}, \nonumber \\
    \tilde{T}_{12} &= \tilde{t}_1G_0\tilde{T}_{22}, \nonumber \\
    \tilde{T}_{21} &= \tilde{t}_2G_0\tilde{T}_{11}, \nonumber \\
    \tilde{T}_{22} &= \tilde{t}_2 + \tilde{t}_2 G_0 \tilde{T}_{12}.
    \label{eq:T_system}
\end{align}
\par In Eq.~\eqref{eq:T_system}, $G_0$ is the propagator of the nucleon in the cluster $\bar{D}_{s1}(2460)$ (the latter, described by a form factor $F_C(\boldsymbol{q})$ in momentum space), and it is given by
\begin{align}
    G_0(\sqrt{s}) &= \int\frac{d^3 q}{(2\pi)^3}\frac{M_N}{\omega_N(\boldsymbol{q})}\frac{1}{2\omega_C(\boldsymbol{q})}\nonumber\\ &\quad\times\frac{F_C(\boldsymbol{q})}{\sqrt{s}-\omega_N(\boldsymbol{q})-\omega_C(\boldsymbol{q})+i\epsilon},
    \label{eq:G_0}
\end{align}
\noindent where $\sqrt{s}$ is the center-of-mass energy of the $n\bar{D}_{s1}$ system, $\omega_N=\sqrt{M_N^2+\boldsymbol{q}^2}$ and $\omega_C=\sqrt{M_C^2+\boldsymbol{q}^2}$. The form factor $F_C(\boldsymbol{q})$  is written as 
\begin{equation}
    F_C(\boldsymbol{q}) = \frac{F(\boldsymbol{q})}{\mathcal{N}}
    \label{eq:FF}
\end{equation}
\begin{align}
    F(\boldsymbol{q})&=\int\limits_{\substack{|\boldsymbol{p}| < q_{\text{max}}, \\ |\boldsymbol{p - q}| < q_{\text{max}}}} \!\! \frac{d^3p}{(2\pi)^3}\frac{1}{M_C - \omega_{\bar{D}^{*}}(\boldsymbol{p})-\omega_{\bar{K}} (\boldsymbol{p})}\nonumber \\ & \qquad \qquad \times\frac{1}{M_C - \omega_{\bar{D}^{*}}(\boldsymbol{p-q})-\omega_{\bar{K}}(\boldsymbol{p-q})}
\end{align}
\noindent with $\mathcal{N}$ being a normalization factor,
\begin{equation}
    \mathcal{N}=F(0)=\int\limits_{|\boldsymbol{p}|<q_{max}}\frac{d^3p}{(2\pi)^3}\left( \frac{1}{M_C - \omega_{\bar{D}^{*}}(\boldsymbol{p})-\omega_{\bar{K}}(\boldsymbol{p})}\right)^2.
\end{equation}
The regulator $q_{max}$ is set to properly reproduce the binding of the cluster, which is attained with $q_{max}=820$~MeV~\cite{Lin:2024hys}. When we perform the integral in $d^3p$ in the form factor, the maximum value allowed for $|\boldsymbol{p}|$ is $2q_{max}$ and $F_C(q)$ becomes zero at this value. 
\par It should be noticed that Eq.~\eqref{eq:G_0} differs from that of former works due to the incorporation of recoil corrections. That is to say, we substitute the $(2M_C)^{-1}$ factor by $(2\omega_C(\boldsymbol{q}))^{-1}$ and, instead of $(q^0 - \omega_N(\boldsymbol{q})+i\epsilon)^{-1}$, we write $(\sqrt{s} - \omega_N(\boldsymbol{q}) - \omega_C(\boldsymbol{q})+i\epsilon)^{-1}$.
\par Equation~\eqref{eq:T_system} can be solved analitically, finding
\begin{align}
\tilde{T}&=
\begin{pmatrix}
	\tilde{T}_{11} & \tilde{T}_{12} \\
	\tilde{T}_{21} & \tilde{T}_{22}
\end{pmatrix}, \nonumber \\
    \tilde{T}_{11} &= \frac{\tilde{t}_1}{1-\tilde{t}_1\tilde{t}_2G_0^2}, \nonumber \\ 
    \tilde{T}_{22} &= \frac{\tilde{t}_2}{1-\tilde{t}_1\tilde{t}_2G_0^2}, \nonumber \\ 
    \tilde{T}_{12} &= \tilde{T}_{21} =  \frac{\tilde{t}_1 \tilde{t}_2G_0}{1-\tilde{t}_1\tilde{t}_2G_0^2}.
\end{align}
\par At this point, we should mention that the vector meson-baryon amplitudes describing the $n\bar{D}^*$ system carry a $\boldsymbol{\epsilon}\cdot\boldsymbol{\epsilon}'$ factor, where $\boldsymbol{\epsilon} \ (\boldsymbol{\epsilon}')$ stands for the spatial part of the polarization vector of the incoming (outgoing) vector mesons (see Ref.~\cite{Roca:2005nm} for more details). This factor, however, is common to all contributions present in the $\tilde{T}$-matrix, since when summing over the polarizations of the internal vector mesons, we have that $\sum\limits_{pol}\epsilon'_i\epsilon''_j\epsilon''_j\epsilon_i=\pmb{\epsilon}\cdot\pmb{\epsilon}^\prime$ and so on. This means that the factor $\pmb{\epsilon}\cdot\pmb{\epsilon}^\prime$ can be factorized in the $\tilde{T}$-matrix, resulting in $\tilde{T}$ being degenerate for spins 3/2 and 1/2.
\par
\begin{figure}[htbp]
  \centering
  \subfigure[]{\includegraphics[scale=0.35]{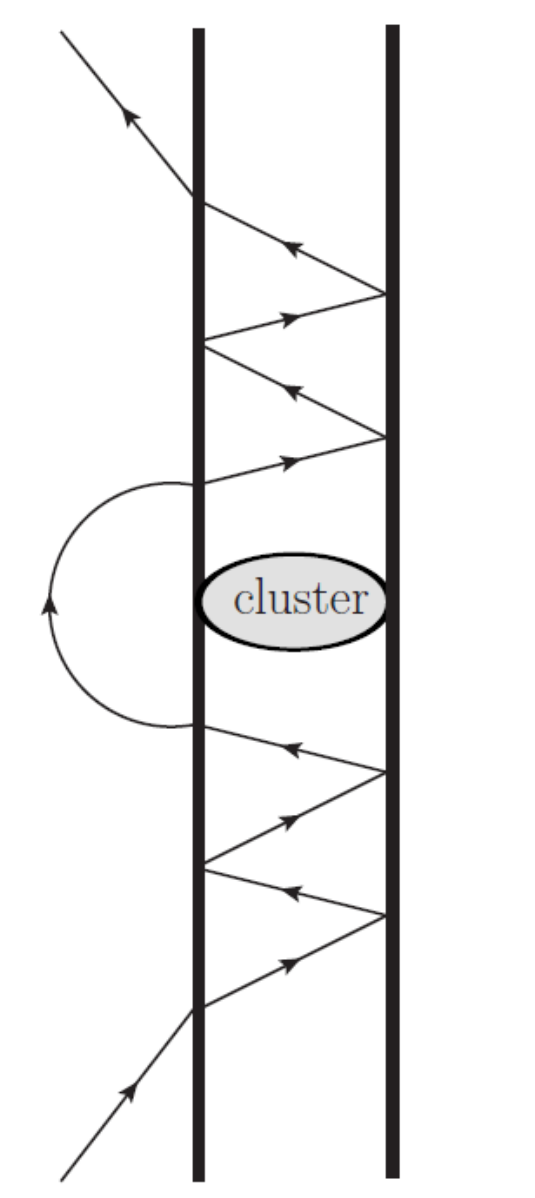}}
  \subfigure[]{\includegraphics[scale=0.385]{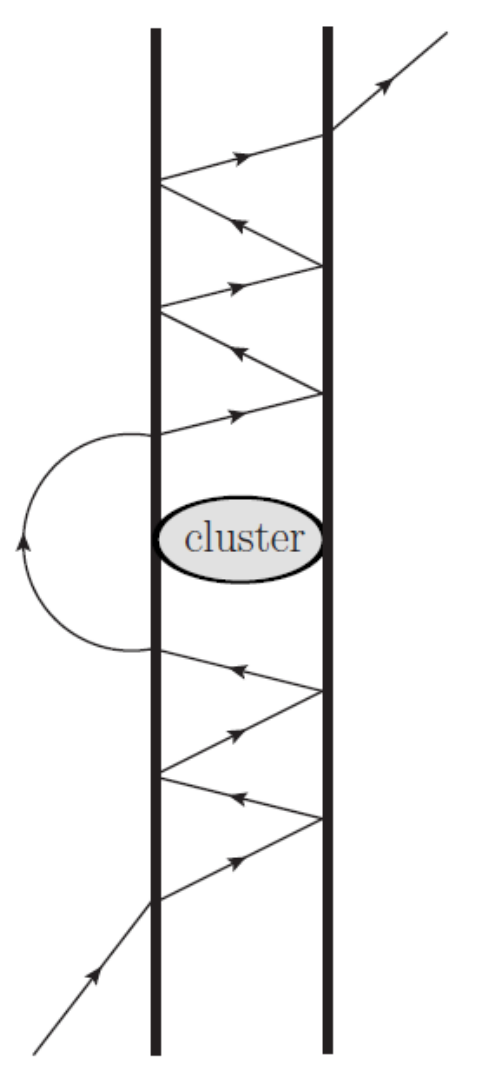}}
  \subfigure[]{\includegraphics[scale=0.345]{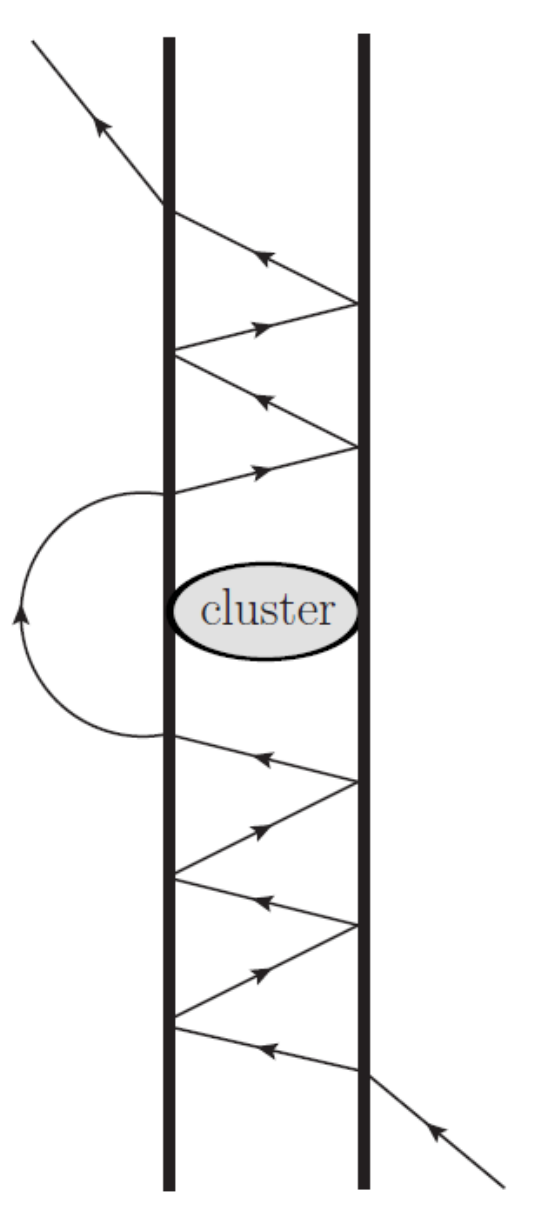}}
  \subfigure[]{\includegraphics[scale=0.395]{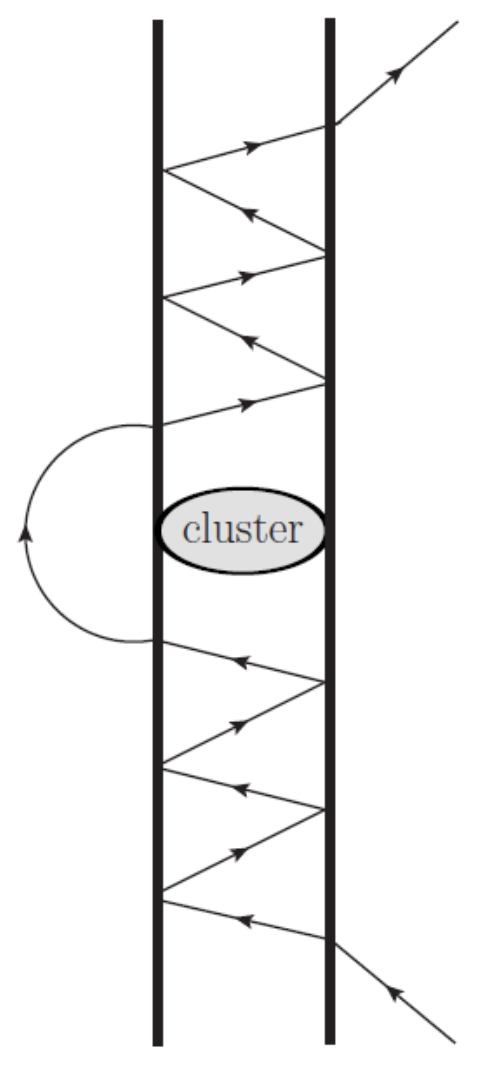}} \\
  \subfigure[]{\includegraphics[scale=0.387]{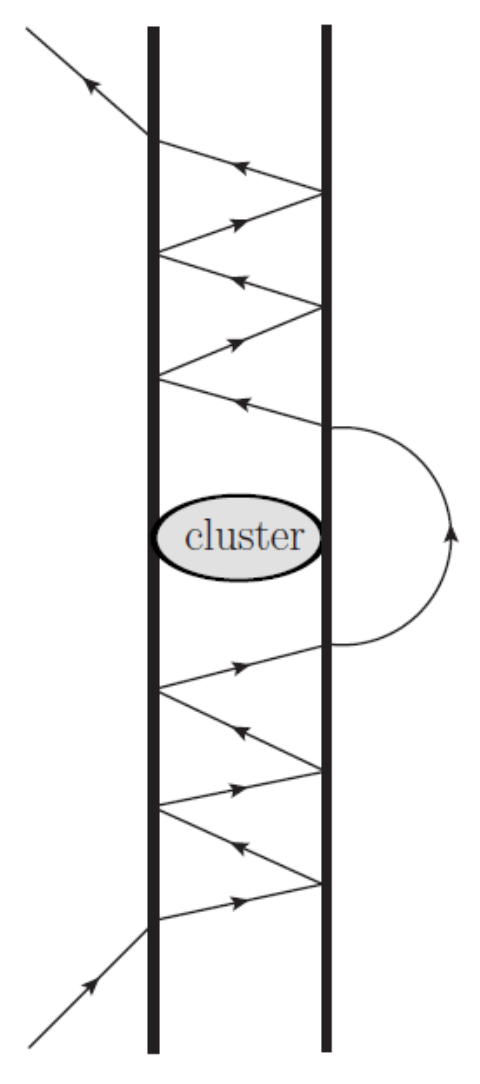}}
  \subfigure[]{\includegraphics[scale=0.395]{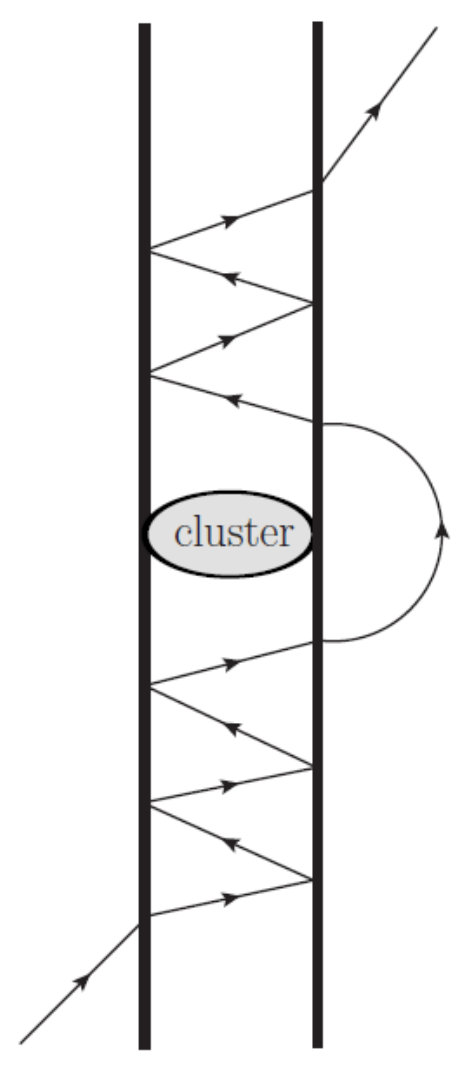}}
  \subfigure[]{\includegraphics[scale=0.348]{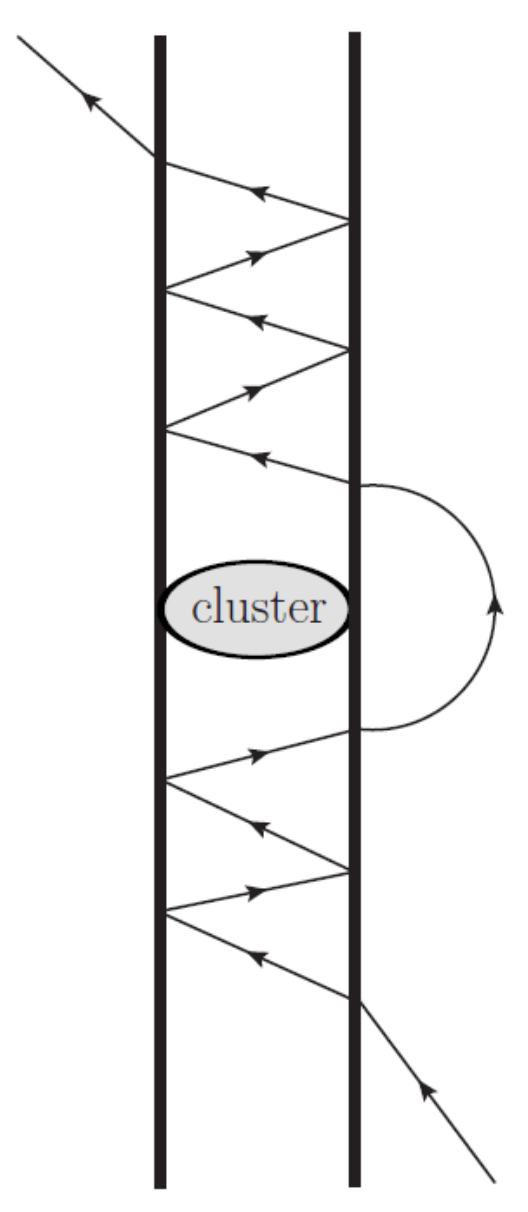}}
  \subfigure[]{\includegraphics[scale=0.348]{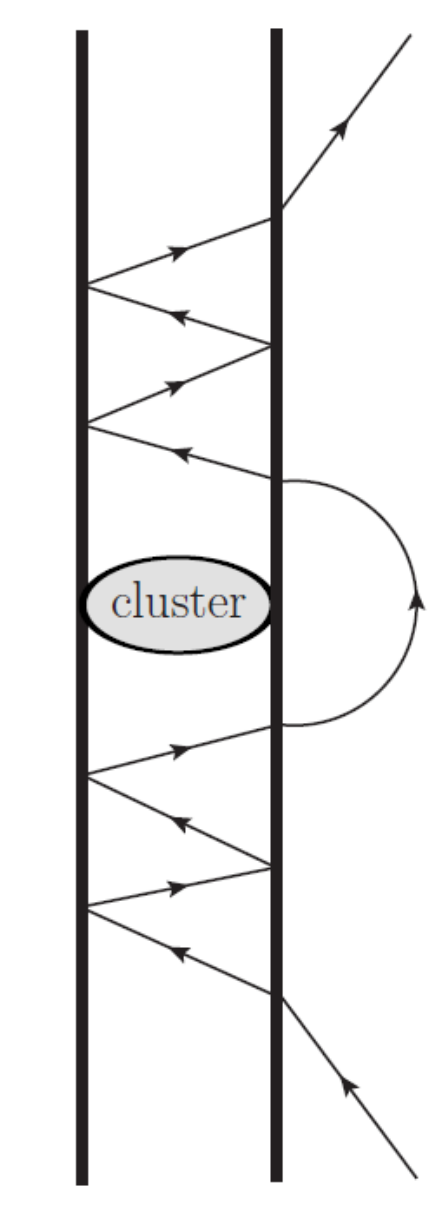}}
  \caption{Feynman diagrams taking into account the elastic propagation of $n$ and the cluster.}
  \label{fig:FCA_diagrams}
\end{figure}

 Next, we consider contributions from the coherent propagation of $n$ and the cluster. For this we have to sum the diagrams shown in Fig.~\ref{fig:FCA_diagrams}. Mathematically, the contributions of the diagrams can be written as: (a) $\tilde{T}_{11}G^{(1)}_{C}\tilde{T}_{11}$, (b) $\tilde{T}_{11}G^{(1)}_{C}\tilde{T}_{12}$, (c) $\tilde{T}_{21}G^{(1)}_{C}\tilde{T}_{11}$, (d) $\tilde{T}_{21}G^{(1)}_{C}\tilde{T}_{12}$, (e) $\tilde{T}_{12}G^{(2)}_{C}\tilde{T}_{21}$, (f) $\tilde{T}_{12}G^{(2)}_{C}\tilde{T}_{22}$, (g) $\tilde{T}_{22}G^{(2)}_{C}\tilde{T}_{21}$, (h) $\tilde{T}_{22}G^{(2)}_{C}\tilde{T}_{22}$.
 
      Here, $G_C^{(1)}$ and $G_C^{(2)}$ represent the neutron--cluster propagation, and since the neutron can interact with either particle 1 or 2 of the cluster, the form factor used in $G_C^{(1)}$ and $G_C^{(2)}$ is different. Following Ref.~\cite{Ikeno:2025bsx}, we have
       \begin{align}
    G^{(i)}_{C}(\sqrt{s})&=\int\frac{d^3 q}{(2\pi)^3}\frac{M_N}{\omega_N(\boldsymbol{q})}\frac{1}{2\omega_C(\boldsymbol{q})}\nonumber\\ &\quad \times\frac{[F^{(i)}_C(\boldsymbol{q})]^2}{\sqrt{s}-\omega_N(\boldsymbol{q})-\omega_C(\boldsymbol{q})+i\epsilon},
    \label{eq:Gcs}
\end{align}
\noindent where $i=1,2$, and the form factors $F^{(i)}_C(\boldsymbol{q})$ can be written in terms of Eq.\eqref{eq:FF} as~\cite{Yamagata11}
\begin{align}
    F^{(1)}_C(\boldsymbol{q}) = F_C\left(\frac{M_{\bar{K}}}{M_{\bar{D}^*}+M_{\bar{K}}}\boldsymbol{q}\right)\nonumber \\
    F^{(2)}_C(\boldsymbol{q}) = F_C\left(\frac{M_{\bar{D}^*}}{M_{\bar{D}^*}+M_{\bar{K}}}\boldsymbol{q}\right),
    \label{eq:FFcs}
\end{align}
with the assumption that the external momenta of the $n$ are small compared to $\boldsymbol{q}$. The terms of Fig.~\ref{fig:FCA_diagrams} can be summed as $\sum\limits_{ij}T'_{ij}$ with $T'_{ij} =\sum\limits_{m}\tilde{T}_{im}G_{c}^{(m)}\tilde{T}_{mj}$. However, we should iterate this $n$-cluster propagation, which is then accomplished via a Lippmann-Schwinger like expression leading finally to $\mathbb{T}_{ij}$, written in matrix form as
\begin{align}
    \mathbb{T} &= \tilde{T} + \tilde{T}G_C \mathbb{T} \nonumber \\
     &=\left[1-\tilde{T}G_C\right]^{-1}\tilde{T},
     \label{eq:T_matrix}
\end{align}
  \noindent where $G_C=\begin{pmatrix}
          G^{(1)}_{C} & 0 \\ 0 & G^{(2)}_{C} 
      \end{pmatrix}$.
    Equation \eqref{eq:T_matrix} can be solved analytically as in Ref.~\cite{Ikeno:2025bsx}, finding 
\begin{align}
T&=\sum_{i,j=1}^2 \mathbb{T}_{ij},\nonumber\\
&=\frac{\tilde{T}_{11}+2\tilde{T}_{12}+\tilde{T}_{22}+(\tilde{T}^2_{12}-\tilde{T}_{11}\tilde{T}_{22})(G_C^{(1)}+G_C^{(2)}) }{1-\tilde{T}_{11}G_C^{(1)}-\tilde{T}_{22}G_C^{(2)} - (\tilde{T}^2_{12}-\tilde{T}_{11}\tilde{T}_{22})G_C^{(1)}G_C^{(2)} }.
\end{align}
\par However this formula can be reduced and written directly in terms of $\tilde{t}_1$, $\tilde{t}_2$, $G_0$, $G_C^{(1)}$, $G_C^{(2)}$ as
\begin{equation}
    T = \frac{\tilde{t}_1+\tilde{t}_2+(2G_0-G_C^{(1)}-G_C^{(2)})\tilde{t}_1\tilde{t}_2}{1-G_C^{(1)}\tilde{t}_1-G_C^{(2)}\tilde{t}_2-(G_0^2-G_C^{(1)}G_C^{(2)})\tilde{t}_1\tilde{t}_2}.
    \label{eq:Alberto_T}
\end{equation}
Note that setting $G^{(1)}_{C}=G^{(2)}_{C}=0$, produces $T=\sum\limits_{i,j=1}^2 \tilde{T}_{ij}$, as expected, finding the result obtained in Ref.~\cite{Encarnacion:2025lyf}. By examining Eq.~\eqref{eq:Alberto_T}, the impact of considering the elastic propagation of the $n$ and $\bar{D}_{s1}$ as a cluster becomes clearer, as we shall see below.

\subsection{The \texorpdfstring{$n\,\bar{D}_{s1}(2536)$}{n Ds1(2536)} system}

\par Following Ref.~\cite{Gamermann:2007fi}, the state $D_{s1}(2536)$ can be considered to be dynamically generated from the $DK^*$ interaction in the $C=1$, $S=1$ sector, with $I=0$ and $J^P=1^+$. In this way,
\begin{align}
    \ket{D_{s1}(2536)} = \frac{1}{\sqrt{2}}\left( \ket{D^{+}K^{*0}} + \ket{D^{0}K^{*+}} \right).
    \label{eq:Ds12536}
\end{align}
Comparing Eqs.~\eqref{eq:Ds12536} and $\eqref{eq:Ds12460}$, to study the $n\bar{D}_{s1}(2536)$ system we just simply need to replace $\bar{D}^* \rightarrow \bar{D}$ and $\bar{K} \rightarrow \bar{K}^*$. By doing this, the two body amplitudes $t_1$ and $t_2$ are now given by
\begin{align}
    t_1(\sqrt{s_{n\bar{D}}}) \equiv t_1 &= \frac{3}{4}t_{n\bar{D}}^{I=1} + \frac{1}{4}t_{n\bar{D}}^{I=0}, \nonumber \\
    t_2(\sqrt{s_{n\bar{K}^*}}) \equiv t_2 &= \frac{3}{4}t_{n\bar{K}^*}^{I=1} + \frac{1}{4}t_{n\bar{K}^*}^{I=0},
\end{align}
with
\begin{align}
    s_{n\bar{D}} &= (p_n + p_{\bar{D}})^2 = m_n^2 + (\xi M_{\bar{D}})^2 + 2\xi M_{\bar{D}} q^0, \nonumber \\
    s_{n\bar{K}^*} &= (p_n + p_{\bar{K}^*})^2 = m_n^2 + (\xi M_{\bar{K}^*})^2 + 2\xi M_{\bar{K}^*} q^0, \nonumber \\
    \xi &= \frac{M_C}{M_{\bar{D}} + M_{\bar{K}^*}}.
\end{align}
Next, the normalized $t_1$ and $t_2$ amplitudes are given by
\begin{equation}
    t_1\rightarrow \tilde{t}_1=\frac{M_C}{M_{\bar{D}}}t_1, \qquad t_2\rightarrow \tilde{t}_2=\frac{M_C}{M_{\bar{K}^*}}t_2,
\end{equation}
with $M_C$ standing now for the mass of $D_{s1}(2536)$. 
\par For this system, the form factor is given now by 
\begin{align}
    F(\boldsymbol{q})&=\int\limits_{\substack{|\boldsymbol{p}| < q_{\text{max}}, \\ |\boldsymbol{p - q}| < q_{\text{max}}}} \!\! \frac{d^3p}{(2\pi)^3}\frac{1}{M_C - \omega_{\bar{D}}(\boldsymbol{p})-\omega_{\bar{K}^*} (\boldsymbol{p})}\nonumber \\ & \qquad \qquad \times\frac{1}{M_C - \omega_{\bar{D}}(\boldsymbol{p-q})-\omega_{\bar{K}^*}(\boldsymbol{p-q})}
\end{align}
\noindent with the normalization
\begin{equation}
    \mathcal{N}=\int\limits_{|\boldsymbol{p}|<q_{max}}\frac{d^3p}{(2\pi)^3}\left( \frac{1}{M_C - \omega_{\bar{D}}(\boldsymbol{p})-\omega_{\bar{K}^*}(\boldsymbol{p})}\right)^2.
\end{equation}
Here we set $q_{max}=1025$~MeV in order to properly reproduce the binding for the $\bar{D}_{s1}(2536)$~\cite{Lin:2024hys}.
\par The last thing to mention are the changes in the form factors in $G^{(1)}_{C}$ and $G^{(2)}_{C}$, which for the $n\bar D_{s1}(2536)$ system are given by $F^{(1)}_C(\boldsymbol{q}) = F_C\!\left(\frac{M_{\bar{K}^*}}{M_{\bar{D}} + M_{\bar{K}^*}}\,\boldsymbol{q}\right)$ and 
$F^{(2)}_C(\boldsymbol{q}) =  F_C\!\left(\frac{M_{\bar{D}}}{M_{\bar{D}} + M_{\bar{K}^*}}\,\boldsymbol{q}\right)$.

\section{ELASTIC UNITARITY}

In quantum mechanics, the scattering amplitude $f^{QM}$ and the two-body $T$-matrix are related via~\cite{MandlShaw}
\begin{equation}
    -\frac{8\pi\sqrt{s}}{2M_N} T^{-1} = (f^{QM})^{-1}.
\end{equation}
For energies near the threshold, we can use the \textit{Effective Range Expansion} (ERE)~\cite{Bethe1949EffectiveRange} and write
\begin{equation}
    -\frac{8\pi\sqrt{s}}{2M_N} T^{-1} = (f^{QM})^{-1} \approx -\frac{1}{a} + \frac{1}{2}r_0 q_{cm}^2 - iq_{cm}, 
    \label{eq:ERE}
\end{equation}
\noindent where $a$ is the scattering length, $r_0$ represents the effective range, and $q_{cm}$ is the modulus of the center-of-mass momentum of the $n\bar{D}_{s1}$ system. Note that it is the factor $-iq_{cm}$ in Eq.~\eqref{eq:ERE} the one that guarantees unitarity. One can prove that the $T$-matrix in Eq.\eqref{eq:Alberto_T} satisfies exactly unitarity. This is done by looking at the linear terms in $q_{cm}$ of $T^{-1}$.
\par In order to see this, we note that the imaginary part of $G$ can be written as 
\begin{align}
    \text{Im}\{G_0\}&=-\frac{2M_N}{8\pi\sqrt{s}}q_{cm}F_C(q_{cm}), \nonumber \\
    \text{Im}\{G^{(i)}_{C}\}&=-\frac{2M_N}{8\pi\sqrt{s}}q_{cm}[F^{(i)}_C(q_{cm})]^2,
    \label{eq:ImG}
\end{align}
\noindent but the form factor behaves as
\begin{equation}
F_C(q) \simeq 1 - \frac{1}{6}\langle r^2 \rangle q_{cm}^2,
\end{equation}
and also similarly for $F^{(i)}_{C}$, with $F_C(0)=F^{(i)}_{C}(0)=1$. Hence, near threshold, we have at linear level in $q_{cm}$
\begin{align}
 \text{Im}\{G_0\}\simeq\text{Im}\{G^{(i)}_{C}\}\simeq -\frac{2M_N}{8\pi\sqrt{s}}q_{cm},
 \label{ImGth}
\end{align}
taking $\sqrt{s}$ at threshold.\par From Eq.~\eqref{eq:Alberto_T}, the inverse of $T$ is given by
\begin{equation}
    T^{-1} = \frac{1-G_C^{(1)}\tilde{t}_1-G_C^{(2)}\tilde{t}_2-(G_0^2-G_C^{(1)}G_C^{(2)})\tilde{t}_1\tilde{t}_2}{\tilde{t}_1+\tilde{t}_2+(2G_0-G_C^{(1)}-G_C^{(2)})\tilde{t}_1\tilde{t}_2}.
    \label{eq:inv_T}
\end{equation}
\par Then, expanding Eq.~\eqref{eq:inv_T} up to linear terms in $q_{cm}$ and using Eq.~\eqref{ImGth}, we satisfy unitarity exactly. In Appendix~\ref{app:Un_G} the reader can see more details.
\par Once the unitarity of the $T$-matrix has been established, the scattering length and effective range are obtained as in Ref.~\cite{Encarnacion:2025lyf}:
\begin{align}
    a&=\frac{2M_N}{8\pi\sqrt{s}}\left. T \right|_{th}, \\
    r_0&=\frac{1}{\mu}\left[\frac{\partial}{\partial\sqrt{s}}\left(-\frac{8\pi\sqrt{s}}{2M_N}T^{-1} + iq_{cm}\right)\right]_{th},
\end{align}
\noindent where $\mu$ is the reduced mass of the three-body system, given by $\mu=M_NM_C/(M_N+M_C)$. As we are close to threshold, $\sqrt{s}=M_N+M_C+q_{cm}^2/2\mu$ and the subscript $th$ means we are calculating at threshold. 

\section{CORRELATION FUNCTION}

\par\  Following Ref.~\cite{Vidana:2023olz}, we can write the correlation function for the $n\bar{D}_{s1}$ system, $C_{n\bar{D}_{s1}}$, in terms of the corresponding $T$-matrix. The correlation function involves the half-off-shell scattering matrix, $\mathbb{T}(\boldsymbol{p},\boldsymbol{q})$, with $\boldsymbol{p}$ an external momentum and $\boldsymbol{q}$ a variable in the loop. Once again we find the form factor $F^{(i)}_C(\boldsymbol{p})F^{(j)}_C(\boldsymbol{q})\simeq F^{(j)}_C(\boldsymbol{q})$ since $\boldsymbol{p}$ is small. Then we must differentiate in the $T$-matrix the terms with the last interaction involving either particle 1 or 2, which leads us to the formula
\begin{align}
C_{n\bar{D}_{s1}}(p) &= 1 + 4\pi \int_0^{\infty} dr\, r^2 S_{12}(r)
\nonumber \\ &\quad \times\left\{ \left| j_0(pr) + T'G' \right|^2
- j_0^2(pr) \right\},
\label{eq:cor_func}
\end{align}
\noindent where $T'G'=(\mathbb{T}_{11} + \mathbb{T}_{12}) G_1(\sqrt{s},r) + (\mathbb{T}_{12} +  \mathbb{T}_{22}) G_2(\sqrt{s},r)$ and $G_1(\sqrt{s},r)$, $G_2(\sqrt{s},r)$ are given by
\begin{align}
    G_{1}(\sqrt{s},r)&=\int\frac{d^3 q}{(2\pi)^3}\frac{M_N}{\omega_N(\boldsymbol{q})}\frac{1}{2\omega_C(\boldsymbol{q})}\nonumber \\ &\quad\times\frac{j_0(qr)F^{(1)}_C(\boldsymbol{q})}{\sqrt{s}-\omega_N(\boldsymbol{q})-\omega_C(\boldsymbol{q})+i\epsilon}, \nonumber \\
    G_{2}(\sqrt{s},r)&=\int\frac{d^3 q}{(2\pi)^3}\frac{M_N}{\omega_N(\boldsymbol{q})}\frac{1}{2\omega_C(\boldsymbol{q})}\nonumber \\ &\quad\times\frac{j_0(qr)F^{(2)}_C(\boldsymbol{q})}{\sqrt{s}-\omega_N(\boldsymbol{q})-\omega_C(\boldsymbol{q})+i\epsilon}.
    \label{eq:G_cor}
\end{align}
\par In Eq.\eqref{eq:cor_func}, $S_{12}(r)$ represents the source function, which gives the probability at given $r$ for the pair production and is parameterized as a Gaussian
\begin{equation}
    S_{12}(r) = \frac{1}{(4\pi R^2)^{3/2}}\exp(-r^2/4R^2),
\end{equation}
with $R$ being a parameter which, for $pp$ collisions, is of the
order of 1~fm, and for heavy-ion collisions $R \sim 5$~fm (we
shall vary $R$ in the range $1{-}2$~fm). The $j_0(x)$ in Eq.~\eqref{eq:G_cor}
refers to the spherical Bessel function, and $p$ to the modulus of the center-of-mass momentum of the $n\bar{D}_{s1}$ system.
The form factors $F^{(1)}_C(\boldsymbol{q})$ and $F^{(2)}_C(\boldsymbol{q})$ in Eq.~\eqref{eq:G_cor}
are those of Eq.~\eqref{eq:FFcs}. We recall that we have considered the approximation
$F^{(i)}_C(\boldsymbol{q})F^{(i)}_C(\boldsymbol{p}) \simeq F^{(i)}_C(\boldsymbol{q})F^{(i)}_C(0) = F^{(i)}_C(\boldsymbol{q})$,
since $|\boldsymbol{p}| \ll |\boldsymbol{q}|$, with $\boldsymbol{p}$
being the three-momentum of the neutron.

\section{NUMERICAL RESULTS}

\subsection{The $n\bar D_{s1}(2460)$ system}

\begin{figure}[htbp]
\begin{center}
\includegraphics[scale=0.5]{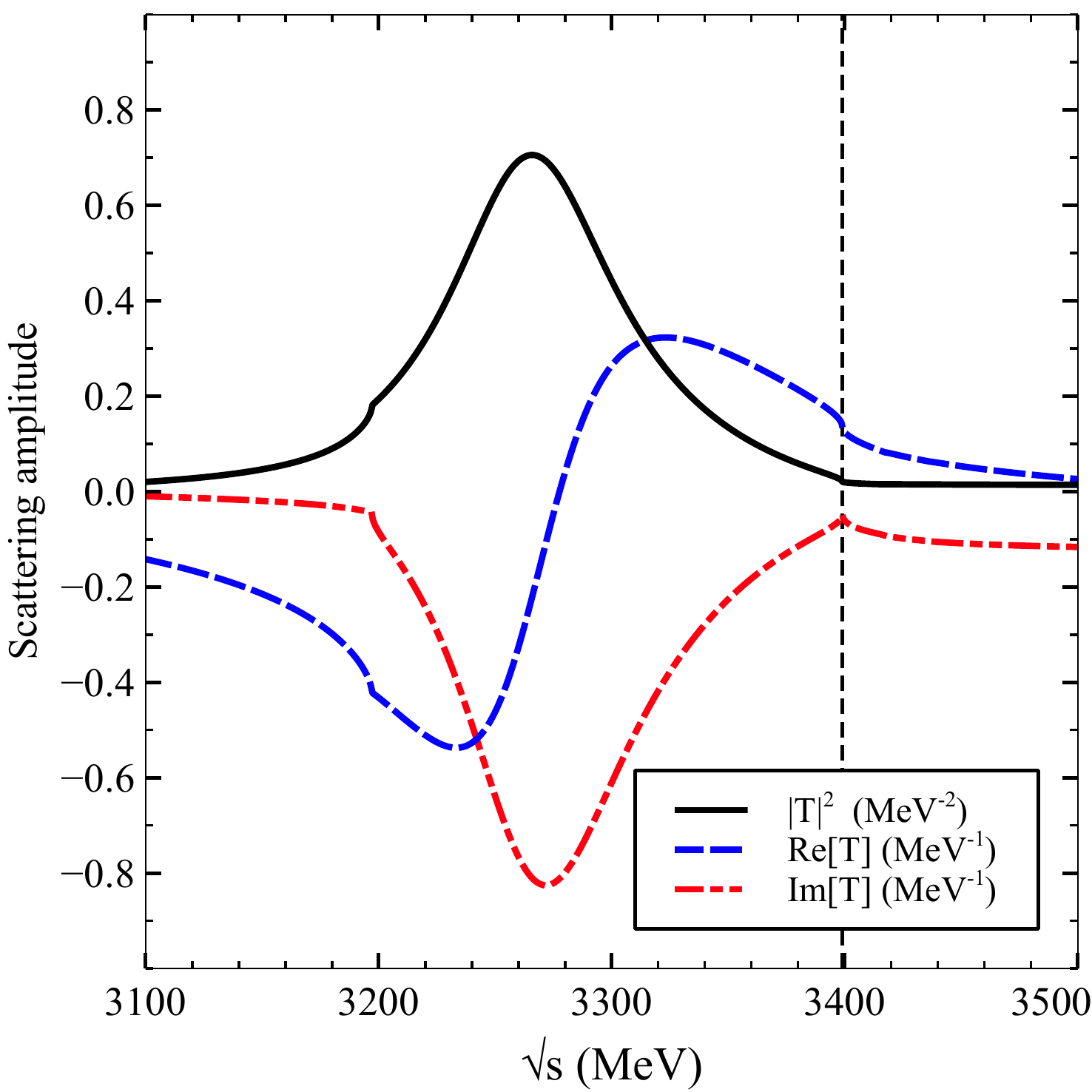}
\caption{Re$[T]$, Im$[T]$ and $|T|^2$ for the $n\bar{D}_{s1}(2460)$ interaction as a function of $\sqrt{s}$.}
\label{fig:amp_2460}
\end{center}
\end{figure}

In Fig.~\ref{fig:amp_2460} we show the scattering amplitude $T$ for the $n\bar D_{s1}(2460)$ system. The solid line represents $|T|^2$, the dashed line the real part of the amplitude, $\text{Re}[T]$, and the dashed dot-dot line the imaginary part, $\text{Im}[T]$. The  vertical dashed line in $\sqrt{s}=3399$~MeV represents the particle + cluster threshold, $(M_C+M_N)$. The peak in $|T|^2$ is located around $3265$~MeV, about $130$~MeV below the $n\bar D_{s1}(2460)$ threshold. Its width is about $90$~MeV. The width originates from the transition of $n\bar{K}$ to $\pi\Sigma$.
\par As we assume the $\bar D_{s1}(2460)$ to be a $\bar K \bar D^*$ molecular state, the attraction stems from the $n\bar{K}$ interaction, responsible for the formation of the $\Lambda(1380)$ and $\Lambda(1420)$ states (replacing the old  $\Lambda(1405)$)~\cite{Oller:2000fj,Jido:2003cb,MartinezTorres:2012yi,Hyodo:2011ur,Oset:2012gi}.

\begin{figure}[htbp]
\begin{center}
\includegraphics[scale=0.5]{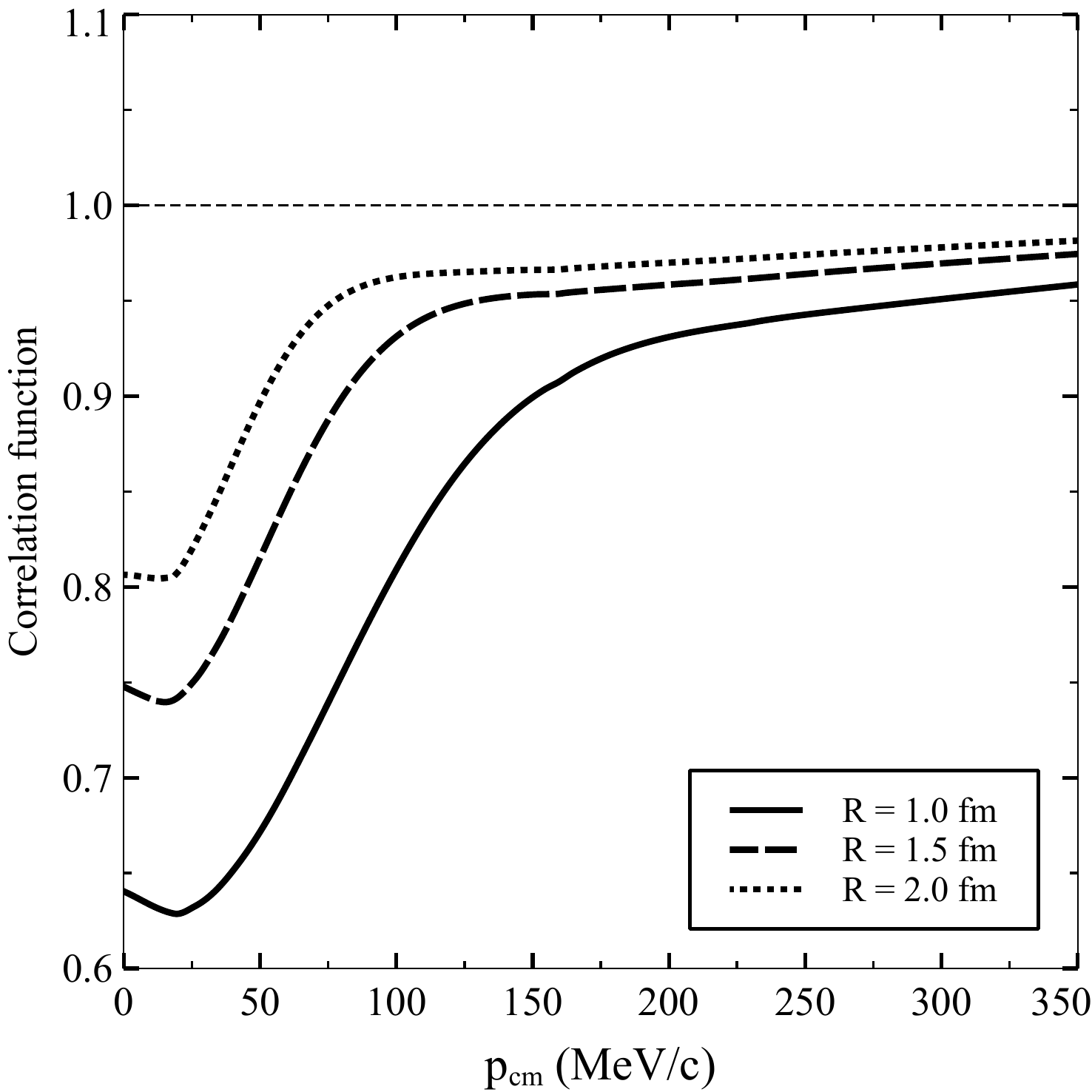}
\caption{Correlation function for the $n\bar{D}_{s1}(2460)$ interaction varying the value of $R$.}
\label{fig:CF_2460}
\end{center}
\end{figure}

The correlation function for the system is given in Fig.~\ref{fig:CF_2460}. The solid line represents the correlation function using $R =1.0$~fm, the dashed line $R = 1.5$~fm and the dotted line $R = 2.0$~fm. We vary $p_{\text{cm}}$ up to $350$~MeV/c, which corresponds to a value of approximately $\sqrt{s}=100$~MeV above the $n\bar{D}_{s1}(2460)$ threshold. From Ref.~\cite{MartinezTorres:2010ax}, we know that for energies of the order of $200$~MeV above the threshold, the FCA is not accurate. That is the reason we do not go further up in momentum. But, as seen in the figure, the relevant structure goes up to $150$~MeV/c, or $25$~MeV excitation energy.
\par The behavior at low momentum in Fig.~\ref{fig:CF_2460} is common when there is a state below threshold~\cite{Ikeno:2023ojl}. For $R = 2.0$~fm the correlation function slowly approaches $1$ as we go further in $p_{\text{cm}}$, and we note that the convergence worsens for smaller R.
\par These results are similar to those obtained for the correlation function of $n\bar D^{*}_{s0}(2317)$ in Ref.~\cite{Ikeno:2025bsx}. The shapes are similar and the strength of the correlation function (diversion from unity) is a bit bigger here than for the $n\bar D^{*}_{s0}(2317)$ case. We also observe that the correlation function falls faster as a function of $p$ than in the case of the $n\bar D^{*}_{s0}(2317)$ system.

The scattering parameters $a$ and $r_0$ for the $n\bar{D}_{s1}(2460)$ system are given by
\begin{align}
a &= (0.59 - i0.21)~\text{fm}, \nonumber \\
r_0 &= (0.65 - i0.16)~\text{fm}.
\end{align}
\par This result for $a$ is very similar to that obtained in Ref.~\cite{Ikeno:2025bsx}, however, the real part of $r_0$ is about one half the one obtained for the $n\bar D^{*}_{s0}(2317)$ case.
\subsection{The $n\bar D_{s1}(2536)$ system}

\begin{figure}[htbp]
\begin{center}
\includegraphics[scale=0.5]{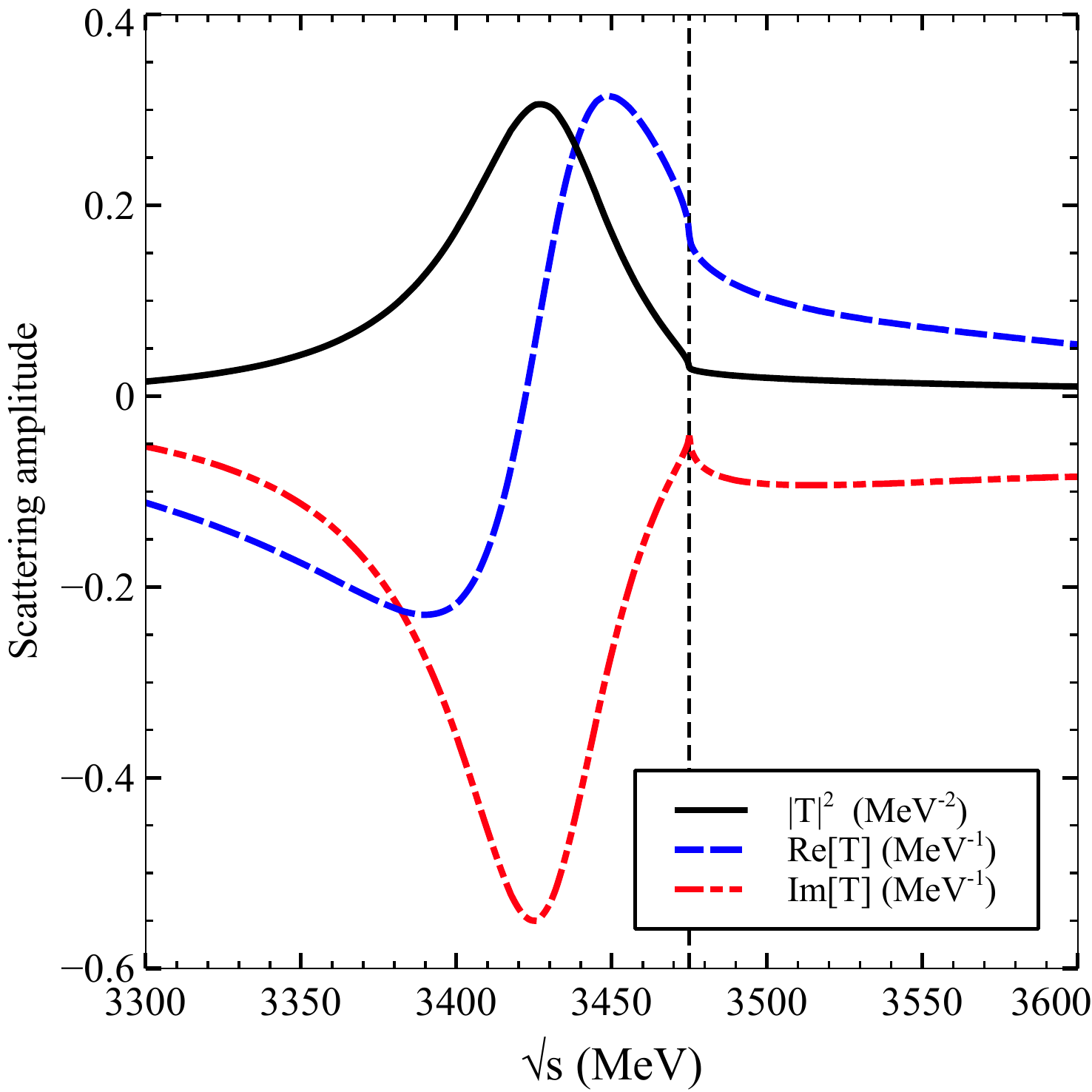}
\caption{Re$[T]$, Im$[T]$ and $|T|^2$ for the $n\bar{D}_{s1}(2536)$ interaction as a function of $\sqrt{s}$.}
\label{fig:amp_2536}
\end{center}
\end{figure}

Next, we show in Fig.~\ref{fig:amp_2536} the behavior of the three-body amplitude $T$ for the $n\bar D_{s1}(2536)$ system. Again, the solid line represents $|T|^2$, the dashed line the real part of the amplitude, $\text{Re}[T]$, and the dashed dot-dot line the imaginary part, $\text{Im}[T]$. The threshold vertical dashed line is now located at $\sqrt{s}=3475$~MeV. Looking at $|T|^2$, there exists a state with mass around $3426$~MeV ($\sim 50$~MeV of biding energy with respect to the $n\bar D_{s1}$ threshold), and a width of $\sim60$~MeV is found. The $N\bar K^*$ interaction here, which provides the attraction for the formation of the the $\Lambda (1800)$ state~\cite{Bethe1949EffectiveRange}, facilitates the formation of a state below threshold. 

\begin{figure}[htbp]
\begin{center}
\includegraphics[scale=0.5]{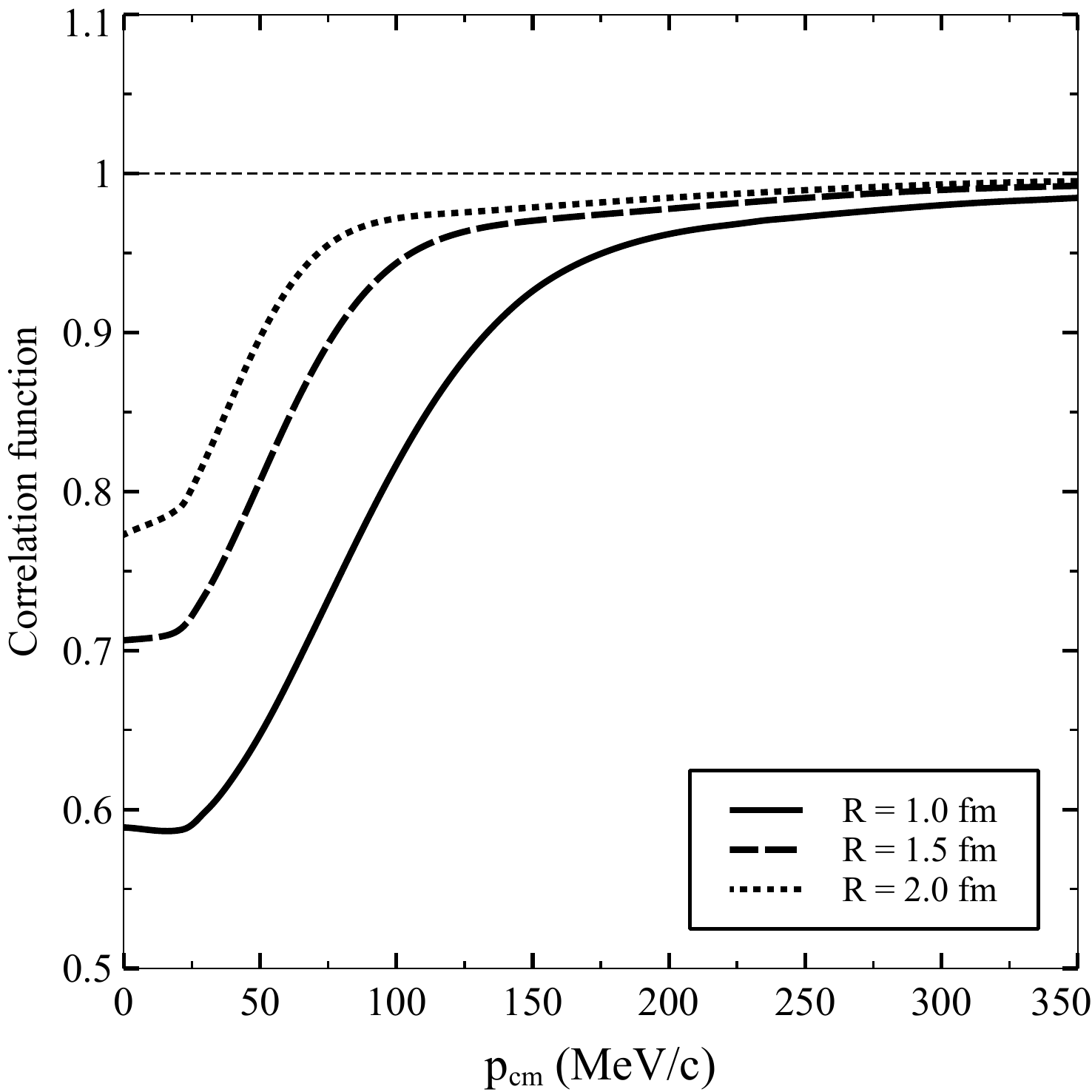}
\caption{Correlation function for the $n\bar{D}_{s1}(2536)$ interaction varying the value of $R$.}
\label{fig:CF_2536}
\end{center}
\end{figure}

Looking at the correlation function shown in Fig.~\ref{fig:CF_2536}, for this system, we find a similar behavior when compared to the $n\bar D_{s1}(2460)$ system. A depletion at low momentum and a growth as we go further in momentum, approaching slowly the unity. The strength of the correlation function is a bit bigger than in the $n\bar D_{s1}(2460)$ case.

The scattering parameters $a$ and $r_0$ for the $n\bar{D}_{s1}(2536)$ system are given by
\begin{align}
a &= (0.71 - i0.18)~\text{fm}, \nonumber \\
r_0 &= (0.16 + i0.32)~\text{fm}.
\end{align}
\par As we can see, the value of $a$ is similar to that of the former system studied, but the value of $r_0$ is quite different, in particular the real part which is about four times smaller. \par At this point it is interesting to recall that the shape and the $R$ dependence of the correlation function that we have found, suggests the presence of a bound state of the system, as is the case here, and which has been discussed before in~\cite{Abreu:2025jqy,Khemchandani:2023xup,Liu:2024nac}.

\section{SUMMARY AND CONCLUSIONS}
We have studied the interaction of a neutron with the states $\bar {D}_{s1}(2460)$ and $\bar {D}_{s1}(2536)$ using an approach based on the Fixed Center Approximation to the Faddeev equations, but improved to satisfy elastic unitarity at threshold. This is essential in order to obtain the scattering length, effective range and correlation function, which have been determined  in this work. We have also assumed that the $D_{s1}(2460)$ and $D_{s1}(2536)$ states are dynamically generated from the $KD^*$ and $K^*D$ interaction, respectively, as comes naturally in the context of the local hidden gauge approach. The system is subject to an overall attraction from the $n \bar K$ or $n \bar K^*$ interaction, which generate the two $\Lambda(1405)$ states or the $\Lambda(1800)$ resonance, respectively, within the context of the chiral unitary approach, which is equivalent to the local hidden gauge approach in this case. This attraction is responsible for the appearance of a resonant state below the $n D_{s1}$ threshold, bound by about $130$~MeV  and with a width of about $90$~MeV in the case of $n \bar {D}_{s1}(2460)$ and $50$~MeV binding with a width of $60$~MeV in the case of  $n \bar {D}_{s1}(2536)$. The unusual width below the threshold is a consequence of the composite structure of the cluster, since $\bar K N$ can go to $\pi \Sigma$. This makes this structure, which one can investigate by looking at the $D^* \pi \Sigma$ invariant mass in experiments, very different than what one would expect if the $D_{s1}$ states corresponded to elementary particles.

In addition to this information, we evaluate the scattering length and effective range for the systems and the correlation functions. The correlation functions have a structure corresponding to a general case when one has a bound state below threshold, which one could confirm once the experimental numbers are available. All this information together should give incentives to measure these correlation functions from where we can obtain information on the nature of the two $D_{s1}$ resonances. Parallely, the study of the three body invariant mass distributions of $D^* \pi \Sigma$ in different experiments should also be encouraged to see the states predicted in the present approach.
\begin{acknowledgments}
This work is partly supported by the
Spanish Ministerio de Economia y Competitividad (MINECO)
and European FEDER funds under Contracts No. FIS2017-
84038-C2-1-P B, PID2020- 112777GB-I00, and by Generalitat Valenciana under contracts PROMETEO/2020/023 and
CIPROM/2023/59. This project has received funding from the
European Union Horizon 2020 research and innovation programme under the program H2020- INFRAIA2018-1, grant
agreement No. 824093 of the STRONG-2020 project. This
work is supported by the Spanish Ministerio de Ciencia e Innovacion (MICINN) under contracts PID2020-112777GB-I00,
PID2023-147458NBC21 and CEX2023-001292-S.
A.M.T, K.P.K, L.M.A, B.A and P.B gratefully acknowledge the partial support provided by the Brazilian agency CNPq (K.P.K: Grants No. 407437/ 2023-1 and
No. 306461/2023-4; A.M.T: Grant No. 304510/2023-8, L.M.A.: Grants No. 400215/2022-5, 308299/2023-0, 402942/2024-8, B.A: Grant No. 987654/2023-1 and 200204/2025-4, P.B: Grant No. 200203/2025-8). Further, A.M.T. thanks Fapesp Grant number 2023/01182-7) and  L. M. A  partly acknowledges the support from CNPq/FAPERJ under the Project INCT-F\'{\i}sica Nuclear e Aplica\c c\~oes (Contract No. 464898/2014-5).
\end{acknowledgments}

\bibliographystyle{apsrev4-2}
\bibliography{references}

\appendix

\section{Two-body amplitudes}
\label{app:amp}

To determine the total amplitude $T$ in Eq.~\eqref{eq:Alberto_T}, we need the
$\tilde{t}_1$ and $\tilde{t}_2$ amplitudes, which, as shown in Eq.~\eqref{eq:2body_t}, depend
on two-body $t$-matrices in specific isospin configurations. In the case of the $n\bar{D}_{s1}(2460)$ system [$n\bar{D}_{s1}(2536)$], we need the $\bar{K}N$ [$\bar{K}^*N$] and $N\bar{D}^*$ [$N\bar{D}$] two-body $t$-matrices in isospin $I=0$ and $I=1$. In the following, we summarize the main aspects for determining such amplitudes, for which the Bethe--Salpeter equation
\begin{equation}
    t = [1 - VG]^{-1}V
    \label{app:BS}
\end{equation}
is solved within a coupled-channel approach. In Eq.~\eqref{app:BS}, $G$ is a two-hadron loop function regularized with a cut-off $q_{\text{max}}$, and the kernel $V$ is obtained from effective
Lagrangians.

\subsection{The $\bar{K}N$ system}

Here we follow the chiral unitary approach of Ref.~\cite{Oset:1997it} with the coupled channels $\bar{K}N$, $\pi \Sigma$, $\eta \Lambda$ and $K\Xi$ for the case of $I=0$ and $\bar{K}N$, $\pi \Sigma$, $\pi \Lambda$, $\eta \Sigma$ and $K\Xi$ for $I=1$. A cut-off $q_\text{max}=630$~MeV is used when calculating $G$. The $V_{ij}^I$ amplitudes describing the $i \rightarrow j$ transition in isospin $I$ and in the center-of-mass frame, with $i,j$ representing the coupled channels, are given by
    \begin{equation}
        V^I_{ij}=-\frac{1}{4f^2}C^I_{ij}(k^0+k'^0),
        \label{app:Vij}
    \end{equation}
    \noindent with $f=1.15f_\pi$ and $f_\pi=93~\text{MeV}$. In Eq.~\eqref{app:Vij}, $k^0$ and $k'^0$ are the energies of the meson present in the initial and final channels, respectively, and they are given by
    \begin{equation}
        k^0 = \frac{s+m_i^2-M_i^2}{2\sqrt{s}}, \quad k'^0 = \frac{s+m_j^2-M_j^2}{2\sqrt{s}}
    \end{equation}
    \noindent with $\sqrt{s}$ representing the center-of-mass energy and $m_{i(j)}$,$M_{i(j)}$ being the masses of the initial (final) mesons and baryons constituting the channel $i (j)$, respectively. The coefficients $C^{I=0}_{ij}$ are given in Table 2 of Ref.~\cite{Oset:1997it}, while $C^{I=1}_{ij}$ can be found in Table 3 of Ref.~\cite{Oset:1997it}.
    
\subsection{The $N\bar{D}^*$ and $N\bar{D}$ interactions}

\par Using the hidden local symmetry approach of Ref.~\cite{Bando:1987br}, the $\bar{D}^{*0}n$, $\bar{D}^{*-}p$ interaction proceeds via the exchange of $\rho,\omega$ in the $t$-channel. This is analogous to the case of the pseudoscalar-baryon chiral Lagrangian, which can be obtained from the exchange of vector mesons in the $t$-channel. Considering the quark content, it is interesting to notice that changing the $\bar{c}$ quark to an $\bar{s}$ quark, $\displaystyle \begin{Bmatrix}\bar{D}^{0*} \\  \bar{D}^{-*} \end{Bmatrix}\rightarrow\begin{Bmatrix}K^{+} \\  K^{0} \end{Bmatrix}$ only the $u$, $d$ quarks, in the form of vector mesons, can be exchanged between the kaon and the nucleon. This means that the $KN$ interaction is completely analogous to that of $N\bar{D}^*$. Thus, we can use the amplitudes obtained in Ref.~\cite{Oset:1997it} for the $KN$ interaction and simply change $M_K\rightarrow M_{\bar{D}^*}$. In this way, in the center-of-mass frame, the amplitudes describing the $N\bar{D}^*\rightarrow  N\bar{D}^*$ interaction are given by
\begin{equation}
        V^I_{ij}=-\frac{1}{4f_\pi^2}L^I(k^0+k'^0).
        \label{app:V_ND}
    \end{equation}
\noindent with $L^{I=0}=0$ and $L^{I=1}=-2$ and $k^0 (k'^0)$ being the energy of the $\bar{D}^*$ meson in the initial (final) state.
\par The same analogy can be used between the $\bar{D}$ and $K$ mesons. Thus, the $N\bar{D}$ interaction is given by Eq.~\eqref{app:V_ND}, with $k^0(k'^0)$ being now the energy of the $\bar{D}$ meson in the initial (final) state.

\subsection{The $\bar{K}^*N$ interaction}
        In Ref.~\cite{Oset:2010tof}, using a coupled channel formalism, the $\bar{K}^*N$ interaction was studied and, for total isospin 0, a state which was identified with $\Lambda(1800)$ was found to be generated. Considering the result found in Ref.~\cite{Oset:2010tof}, here we follow Ref.~\cite{Encarnacion:2025lyf} where the $t$-matrix describing the $\bar{K}^*N\rightarrow \bar{K}^*N$ transition in total isospin 0 is parameterized for energies close to that of $\Lambda(1800)$  as
        \begin{equation}
            T^{I=0}_{\bar{K}^*N}(\sqrt{s})= \frac{g^2_{\bar{K}^*N}}{\sqrt{s}-M_{\Lambda^*}+i\Gamma_{\Lambda^*}/2},
        \end{equation}
        \noindent where $M_{\Lambda^*}$ is the mass of the $\Lambda(1800)$, $\Gamma_{\Lambda^*}$ its width and $g_{\bar{K}^*N}=3.65$, which is an average between the values considered in Ref.~\cite{Garzon:2012np} ($g_{\bar{K}^*N}=4$) and Ref.~\cite{Oset:2010tof} ($g_{\bar{K}^*N}=3.3$).
\par In the case of total isospin 1, the coupled channels $\bar{K}^*N, \ \rho \Lambda, \ \rho\Sigma, \ \omega\Sigma, \ K^*\Xi, \ \phi\Sigma$ were considered in Ref.~\cite{Oset:2010tof} and the corresponding amplitudes are given by 
        \begin{equation}
        V^{I=1}_{ij}=-\frac{1}{4f^2}C^{I=1}_{ij}(k^0+k'^0)\boldsymbol{\epsilon}\cdot\boldsymbol{\epsilon}'.
        \label{app:V_KN}
    \end{equation}
In Eq.~\eqref{app:V_KN}, $\boldsymbol{\epsilon} \ (\boldsymbol{\epsilon}')$ represents the spatial part of the polarization vector related to the vector meson in the inital (final) state. The $C^{I=1}_{ij}$ coefficients can be found in Table 9 of Ref.~\cite{Oset:2010tof}. When solving Eq.~\eqref{app:BS}, we have taken into account the finite width of the $\rho$ and $K^*$ vector mesons. We do this by convoluting $G$ with the corresponding spectral distribution, as done in Ref.~\cite{Geng:2006yb}:
\begin{align}
&\tilde{G}(\sqrt{s}, M, m) = \frac{1}{C} \int_{(M_V - 2\Gamma_V)^2}^{(M_V + 2\Gamma_V)^2} d\tilde{m}^2 \  G(\sqrt{s}, \tilde{m}, m) \nonumber \\
&\qquad \times \left( -\frac{1}{\pi} \right) \text{Im} \left\{ \frac{1}{\tilde{m}^2 - M_V^2 + i M_V \Gamma_V} \right\},
\end{align}
with
\begin{equation}
C = \int_{(M_V - 2\Gamma_V)^2}^{(M_V + 2\Gamma_V)^2} d\tilde{m}^2 \left( -\frac{1}{\pi} \right) 
\text{Im} \left\{ \frac{1}{\tilde{m}^2 - M_V^2 + i M_V \Gamma_V} \right\}.
\end{equation}

\section{Unitarity}
\label{app:Un_G}
We can write the amplitude $T$ in terms of the two-body interactions $\tilde{t}_1$ and $\tilde{t}_2$ as in Eq.~\eqref{eq:Alberto_T}. The inverse of Eq.~\eqref{eq:Alberto_T} is given by 
\begin{equation}
    T^{-1} = \frac{1-G^{(1)}_{C}\tilde{t}_1-G^{(2)}_{C}\tilde{t}_2-(G_0^2-G^{(1)}_{C}G^{(2)}_{C})\tilde{t}_1\tilde{t}_2}{\tilde{t}_1+\tilde{t}_2+(2G_0-G^{(1)}_{C}-G^{(2)}_{C})\tilde{t}_1\tilde{t}_2}.
    \label{eq:AppUn_T}
\end{equation}
\par Writing explicitly the real and imaginary parts of $G_0$, $G^{(1)}_{C}$, $G^{(2)}_{C}$, we have
\begin{align}
G_0^2&=(\text{Re}\{G_0\})^2-(\text{Im}\{G_0\})^2+2i\text{Re}\{G_0\}\text{Im}\{G_0\}, \nonumber \\
G^{(1)}_{C}&G^{(2)}_{C}=\text{Re}\{G^{(1)}_{C}\}\text{Re}\{G^{(2)}_{C}\} - \text{Im}\{G^{(1)}_{C}\}\text{Im}\{G^{(2)}_{C}\}\nonumber \\ &+i(\text{Re}\{G^{(1)}_{C}\}\text{Im}\{G^{(2)}_{C}\} + \text{Re}\{G^{(2)}_{C}\}\text{Im}\{G^{(1)}_{C}\}).
\end{align}
\par The imaginary part of $G_0$, $G^{(1)}_{C}$, $G^{(2)}_{C}$ can be written as in Eq.~\eqref{eq:ImG}.  We are interested in unitarity near the threshold. Then, up to linear order terms in $q_{cm}$
\begin{align}
    \text{Im}\{G_0\} \simeq \text{Im}\{G^{(i)}_{C}\}\simeq -\frac{2M_N}{8\pi\sqrt{s}}q_{cm},
\label{ImG0appen}
\end{align}
where $\sqrt{s}$ is taken at threshold.\par As mentioned before, it is the $-iq_{cm}$ term in Eq.\eqref{eq:ERE} that guarantees unitarity. Hence, using Eq.~\eqref{ImG0appen} and keeping terms up to those linear in $q_{cm}$, we expand Eq.~\eqref{eq:AppUn_T}  and retain only linear terms in $q_{cm}$, and also multiply the result by the factor $-8\pi\sqrt{s}/2M_N$. With this, we have, understanding that only linear terms in $q_{cm}$ for $\left.T^{-1}\right|_\text{lin}$ are kept,
\begin{align}
&\left.\frac{-8\pi\sqrt{s}}{2M_N}T^{-1}\right|_\text{lin}\simeq -\frac{8\pi\sqrt{s}}{2M_N}\frac{1}{\mathcal{D}}\Big(i\frac{2M_N}{8\pi\sqrt{s}}q_{cm} \tilde{t}_1+i\frac{2M_N}{8\pi\sqrt{s}}q_{cm} \tilde{t}_2 \nonumber \\
\quad &+i\frac{2M_N}{8\pi\sqrt{s}}q_{cm} ( 2\text{Re}\{G_0\}-\text{Re}\{G^{(1)}_{C}\}-\text{Re}\{G^{(2)}_{C}\} )\tilde{t}_1\tilde{t}_2\Big) ,
\end{align}
where $\mathcal D = \tilde{t}_1+\tilde{t}_2+( 2\text{Re}\{G_0\}-\text{Re}\{G^{(1)}_{C}\}-\text{Re}\{G^{(2)}_{C}\} )\tilde{t}_1\tilde{t}_2$. Hence, we write
\begin{align}
&\left.\frac{-8\pi\sqrt{s}}{2M_N}T^{-1}\right|_\text{lin}\simeq -\frac{8\pi\sqrt{s}}{2M_N}i\frac{2M_N}{8\pi\sqrt{s}}q_{cm}\nonumber\\ 
&\quad \times\frac{\tilde{t}_1+\tilde{t}_2+( 2\text{Re}\{G_0\}-\text{Re}\{G^{(1)}_{C}\}-\text{Re}\{G^{(2)}_{C}\} )\tilde{t}_1\tilde{t}_2 }{\tilde{t}_1+\tilde{t}_2+( 2\text{Re}\{G_0\}-\text{Re}\{G^{(1)}_{C}\}-\text{Re}\{G^{(2)}_{C}\} )\tilde{t}_1\tilde{t}_2}.
\end{align}
Leading, finally, to
\begin{equation}
-\left.\frac{8\pi\sqrt{s}}{2M_N}T^{-1}\right|_\text{lin}=-iq_{cm},
\end{equation}
as it should be.
\par With this, we prove that the $T$-matrix satisfies exactly the elastic unitarity.

\end{document}